\documentclass[a4paper,12pt]{article}
\usepackage[utf8]{inputenc}
\usepackage[T2A]{fontenc}
\usepackage{indentfirst}
\usepackage{amsmath}
\usepackage{amsthm}
\usepackage{amsfonts}
\usepackage{amssymb}
\usepackage{graphicx}
\usepackage{setspace}
\usepackage{array}
\usepackage{float}
\restylefloat{table}
\usepackage{url}
\usepackage{multirow}
\usepackage{multicol}
\usepackage{enumitem}
\usepackage[title]{appendix}
\usepackage{titling}
\usepackage{longtable}

\usepackage[caption=false]{subfig}
\usepackage{amsbib}
\usepackage{geometry}
\newcommand*{\ket}[1]{| #1 \rangle}
\newcommand*{\bs}[1]{\boldmath #1}

\begin{document}
\large

\par\vspace{-60pt}
\begin{center}
	
	{\large{\bfseries  {Quantum-enhanced symmetric cryptanalysis for S-AES}}} 
	\\[8pt]
	\small{\textbf{Alexey Moiseevskiy}}
	
	\small{\textit{
		Infotecs Scientific Research and Advanced Developments Centre
	}}
\\[8pt]
\end{center}

\begin{abstract}
Advanced Encryption Standard is one of the most widely used and important symmetric ciphers for today. It well known, that it can be subjected to the quantum Grover's attack that twice reduces its key strength. But full AES attack requires hundreds of qubits and circuit depth of thousands, that makes impossible not only experimental research but also numerical simulations of this algorithm. 
Here we present an algorithm for optimized Grover's attack on downscaled Simplified-AES cipher. Besides full attack we present several approaches that allows to reduce number of required qubits if some nibbles of the key are known as a result of side-channel attack. For 16-bit S-AES the proposed attack requires 23 qubits in general case and 19, 15 or 11 if 4, 8 or 12 bits were leaked in specific configuration. Comparing to previously known 32-qubits algorithm this approach potentially allows to run the attack on today’s NISQ-devices and perform numerical simulations with GPU, that may be useful for further research of problem-specific error mitigation and error correction techniques.
\end{abstract}
%
%
%
%
\section{Introduction}

Hybrid quantum algorithms have attracted increasing interest in recent years. Being able to solve
problems more efficiently than classical approaches, they retain acceptable requirements for the depth of
the quantum circuit and the number of qubits used. That increases the stability of the algorithm against
noise and makes it easier to use error mitigation methods. As a result, experimental execution of these
algorithms on NISQ-devices becomes possible, and, no less important, simulation of these algorithms can
be performed on classical computers.
A recent publication \cite{ref1} is a striking example of the effective use of a hybrid approach at the field of of quantum cryptanalysis. In this paper, the original problem is initially subjected to the strong classical processing, including dimensionality reduction, and then the quantum computer is used
only to solve the core of the NP-hard problem on the reduced space with a variational algorithm. As a
consequence, it is possible to achieve sublinear complexity in terms of the number of qubits required for
RSA attack. And although the algorithm complexity is still questionable, the result of resources savings
is very significant, because today the required number of qubits remains the strongest limitation for practical quantum computers usage.
Thus, since the original methods of quantum cryptanalysis based on the Grover, Shor and Simon
algorithms require significantly more qubits than are available in today’s NISQ-devices, it is most likely
that the first practical quantum cryptanalysis tools will be based on a hybrid approach.

%
%

\section{S-AES algorithm}

\begin{figure}
	\centering
	\includegraphics[width=0.5\linewidth]{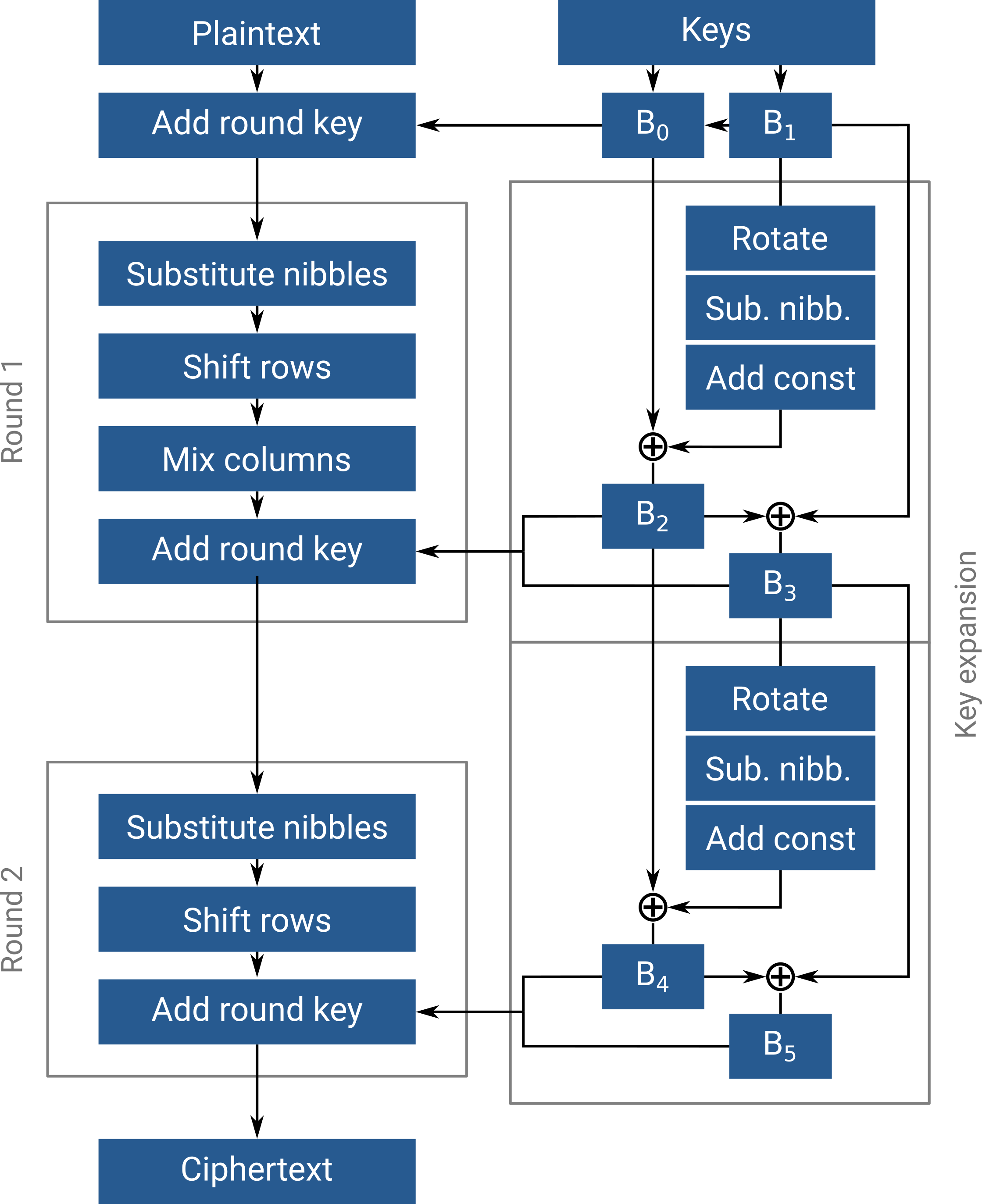}
	\caption{S-AES algorithm. Input: 16-bit text, 16-bit key. Output: 16-bit ciphertext.}
	\label{fig:S-AES}
\end{figure}

S-AES is a simplified version of AES block cipher \cite{saes}. It consists of two rounds and works with 16-bit key and text. Similar to AES there are four basic operations: \textit{Add round key}, \textit{Substitute nibbles}, \textit{Shift rows} and \textit{Mix columns}. The difference is that in S-AES elementary unit is reduced from 1 byte to 1 nibble that stands for 4 bits. Also substitution operation also known as S-box operates with 4 bits instead of 8. Algorithm block-scheme is given at figure \ref{fig:S-AES}.  

\subsection{Grover's search}  

The problem to solve is to recover the key from given plaintext and ciphertext. In the classical approach, for symmetric cryptography it is assumed that in general case there is no more efficient way than sequential enumeration or brute force attack. As a result the key strength grows exponentially with its length. For 16-bit in the worst case $2^{16}$ keys will be enumerated. But for quantum computer it's possible to use Grover's search that provides a quadratic speedup for the brute force attack \cite{Grover}. The main requirement for Grover's search is the need to efficiently implement a quantum program that checks whether the given input is a correct solution to the search problem. Such a program, capable of performing a phase flip of a given quantum state, if the state corresponds to the correct problem answer, is called an oracle. The oracle contains all the information about the problem being solved. The remaining part of Grover's algorithm, called Grover's operator, performs the amplitude amplification of the computational basis state marked by the oracle and does not depend on the problem being solved. After $\left \lfloor{\frac{\pi}{4} \sqrt{N/M}}\right \rfloor $ iterations, where $N$ stands for the number of elements in the search space and $M$ is the number of correct answers, the amplitudes of all input quantum state components, except for the component encoding the correct answer, are practically suppressed, and as a result, the correct answer can be read from the quantum register via measurements in the numerical basis.

In the case of quantum cryptanalysis it is necessary to create a quantum circuit that will encrypt the planetext with a given key and compare the result with the given ciphertext. For this it's necessary to implement all S-AES subroutines as quantum circuits. And the main goal is to reduce the requirements of these schemes to the number of qubits used. It is better if the circuit depth is also reduced as much as possible, but the number of qubits used remains the most restrictive since the circuit depth limitation can be reduced using error mitigation and correction techniques. For today the best known S-AES attack requires 32 qubits which makes it impossible to simulate it on a 40GB GPU. If we can reduce the requirements to 29 qubits, then it will be possible to conduct simulations on the GPU, which will have a significant impact on further research.

\subsection{Quantum circuits}  

\begin{figure*}
	\centering
	\subfloat[Quantum S-box]
	{
		\includegraphics[width=0.49\linewidth]{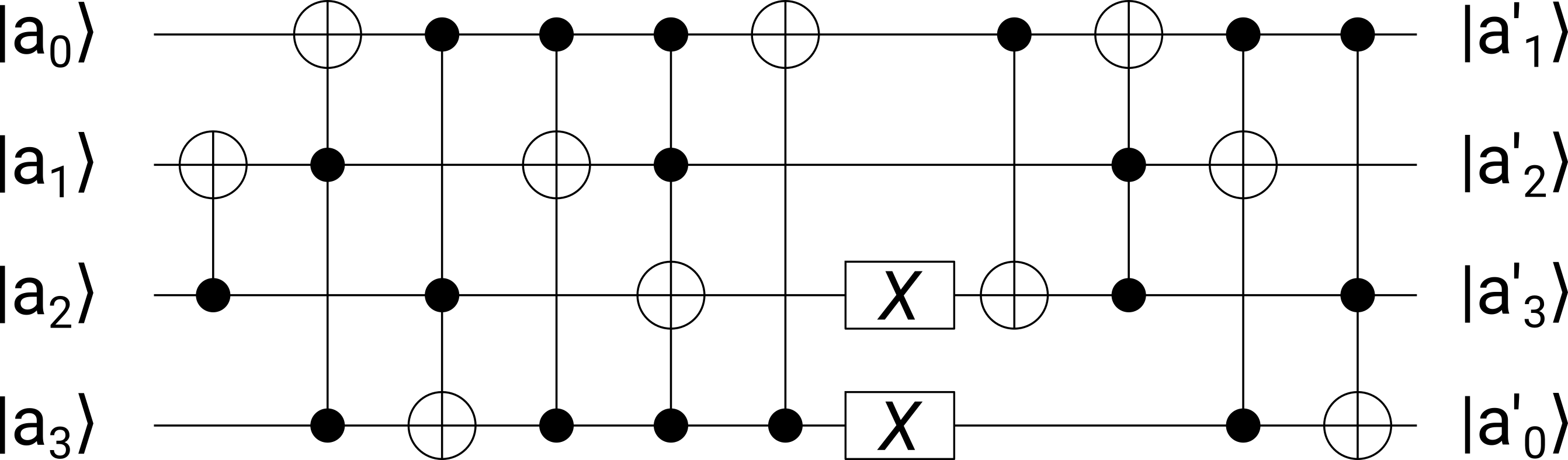}
		\label{fig:Sbox}
	}
	\subfloat[Quantum Mix Columns]
	{
		\includegraphics[width=0.29\linewidth]{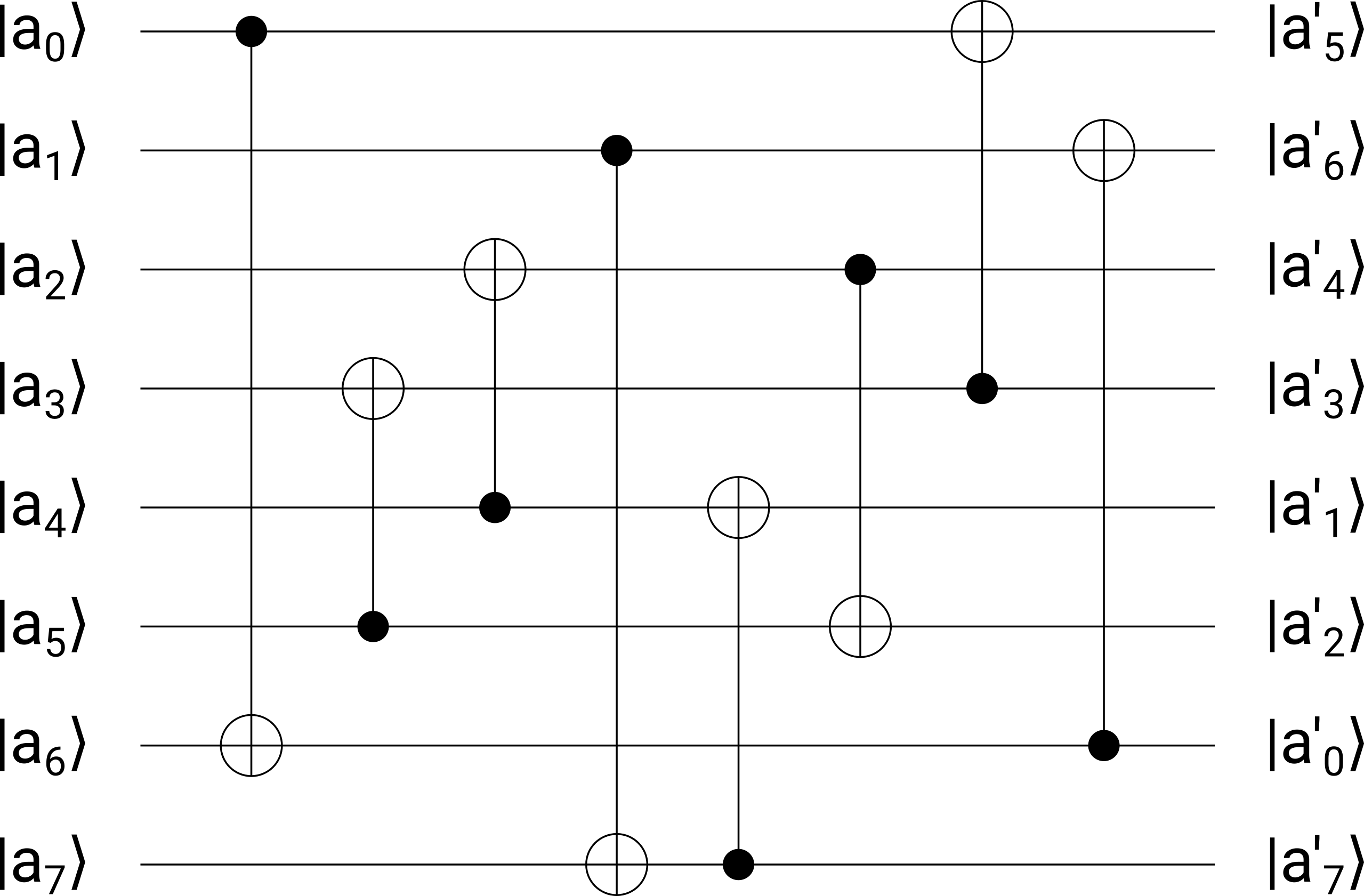}
		\label{fig:MC}
	}
	\caption{Quantum implementation of the 1-nibble S-AES S-box and 2-nibbles Mix Column transformations given in \cite{Grover on Simplified AES}. A big advantage is that for both cases no ancilla qubits are required. CNOT, Toffoli and Pauli-X gates are used.}
	\label{fig:quantumCircuits}
\end{figure*}

Like in the AES the S-AES S-box is non-linear revertible map from nibbles to nibbles. In classical case the S-box can be easily implemented as a map table, but quantum computer has no free memory to store it. Without diving deep into the algebra behind this map, the best known implementation of the S-box in the form of a quantum circuit is presented on figure \ref{fig:Sbox}.

The Mix Column transformation can be represented similarly. Behind the polynomial-algebraic representation, this transformation can be defined as a simple bitwise algorithm for an eight-bit array. 

\begin{equation}
	\begin{aligned}
		a_0' &= a_0 \oplus a_6 &
		a_4' &= a_4 \oplus a_2 \\
		a_1' &= a_1 \oplus a_4 \oplus a_7 &		
		a_5' &= a_5 \oplus a_3 \oplus a_0 \\
		a_2' &= a_2 \oplus a_4 \oplus a_5 &		
		a_6' &= a_6 \oplus a_1 \oplus a_0 \\
		a_3' &= a_3 \oplus a_5 &
		a_7' &= a_7 \oplus a_1 
	\end{aligned}
	\label{eq:MC}
\end{equation}
Apparently optimal quantum circuit for Mix Columns can be taken from this representation. It's shown at figure \ref{fig:MC}.

The Shift Rows transformation simply swaps the second and the fourth nibble for 4-nibble text. For quantum computer it is equivalent to the qubit indexes switch and in terms of quantum operations can be done for free. 

The Add Key operation in classic is just a bitwise xor between the key and the text. In quantum computing if both key and text are stored in qubits this is equivalent to the CNOT operation. But in case of a quantum attack, it often becomes necessary to produce xor of classical and quantum data. Then it can be implemented by dynamically changing the quantum circuit with the addition of Pauli-X operators to the corresponding positions of qubits, depending on the classical data. Obviously, the result of such an operation is quantum data and is stored in a qubit.

The key expansion subroutine also includes Rotate Nibbles and Add Round Constant functions. Add Round Constant acts just like a xor with classical data. And Rotate Nibbles acts on the 2-nibble round key by simply swapping the nibbles.

\begin{figure*}
	\centering
	\subfloat[Quantum circuit for one S-AES round]
	{
		\includegraphics[width=0.33\linewidth]{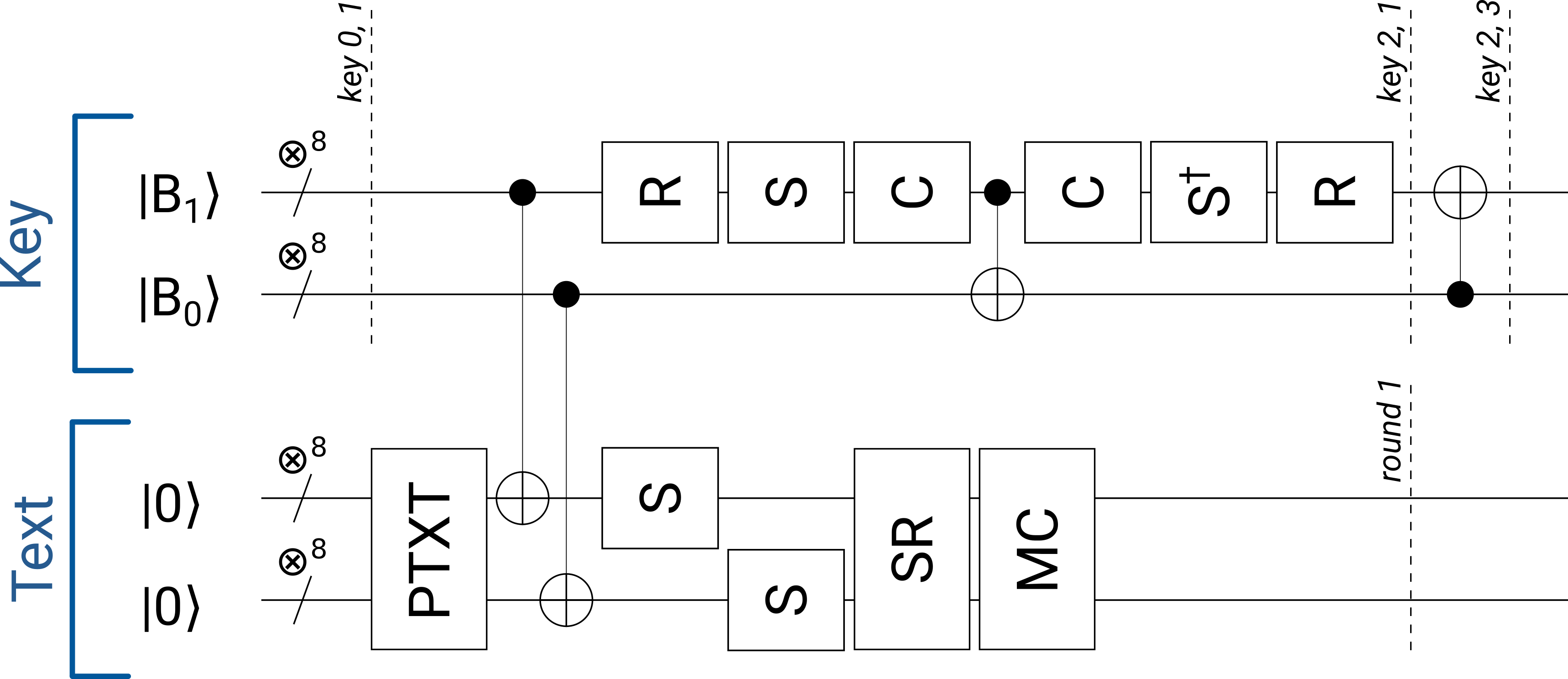}
		\label{fig:Round}
	}
	\subfloat[Quantum circuit for one  iteration of S-AES attack]
	{
		\includegraphics[width=0.6\linewidth]{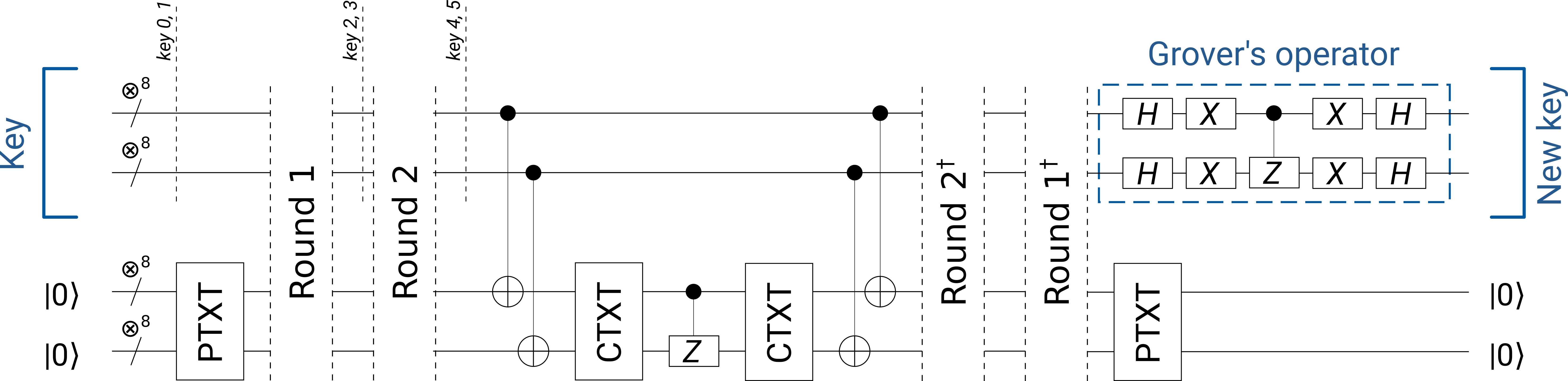}
		\label{fig:Iteration}
	}
	\caption{Quantum circuits for S-AES Grover's attack. \ref{fig:Round} shows one round of the oracle that follows the algorithm from figure \ref{fig:S-AES}. Here the R stands for Rotate Nibble, S is the S-box,  $S^\dagger$ - inverted S-box, C - Add Round Constant, SR - Shift Rows, MC - Mix Columns. PTXT and CTXT add the plaintext and the ciphertext to the register with Pauli-X gates. \ref{fig:Iteration} shows one iteration of the Grover's attack. Inversed rounds are necessary to restore the initial key state after oracle's action. Grover's phase estimation and amplitude amplification operator is shown in blue dashed frame.}
	\label{fig:Oracle}
\end{figure*}

Therefore elementary Grover's oracle for S-AES attack can be constructed us shown at figure \ref{fig:Oracle}. The key idea is to generate the ciphertext for given key state in the text qubit register and then xor it with known ciphertext in such a way that in case of a deterministic match the state $\ket{1}$ is formed for all qubits. Then for an entangled state the component $\ket{11...1}$ in the text register will correspond the correct key component in the key register. Following Controlled-Z (CZ) gate will flip phase of this component which is the main goal of the oracle. Finally, inverted S-AES algorithm will restore the initial round state in the key register and then Grover's operator can be applied again to amplify the amplitude of flipped-phase component that corresponds to the correct key.

\section{Leak-based optimized attack}

The attack described above requires 32 qubits. This is small enough to be able to talk about the experimental implementation of at least one iteration of the algorithm, the depth of the quantum circuit of which can be on the order of a hundred layers, depending on the hardware-implemented gates. But simulation of 32 qubits algorithm, even with a simulator based on noise gates, requires more than 40 GB of RAM, which makes it practically impossible to simulate the attack algorithm on a GPU. 

\subsection{$\boldsymbol{B_1}$-leak 24-qubits exact attack}

The possible way to reduce the number of qubits required is to consider the case where some of the key bits are leaked due to a side channel attack. As you can see in the figure \ref{fig:S-AES} and figure \ref{fig:Round}, in S-AES the 16-bit round keys are divided into two 8-bit segments $B$, as well as the text is divided into two 8-bit segments $N$. Let's consider a simple case where key $B_1$ was leaked. For this, firstly, define the algebraic expressions of the keys: 
\begin{equation}
\begin{aligned}
B_0 &= B_0 & B_3 &= B_1 B_2 = C_0 B_0 B_1 B_1^{SR} \\
B_1 &= B_1 & B_4 &= C_1 B_2 {B_3}^{SR} = C_0 C_1 B_0 B_1^{SR} [C_0 B_0 B_1 B_1^{SR}]^{SR} \\
B_2 &= C_0 B_0 B_1^{SR} & B_5 &= B_3 B_4 =  C_1 B_1 B_2 B_2 {B_3}^{SR}= C_1 B_1 {B_3}^{SR}  
\end{aligned}
\label{eq:keys}
\end{equation}
Here $C_0$ and $C_1$ are round constants, superscripts S and R shows that the argument was subjected to the Substitute and Rotate Nibbles operations. The question here is whether it is possible to use additional data to modify the algorithm and unravel some of the qubits from the rest of the register. 

\begin{figure}
	\centering
	\includegraphics[width=0.7\linewidth]{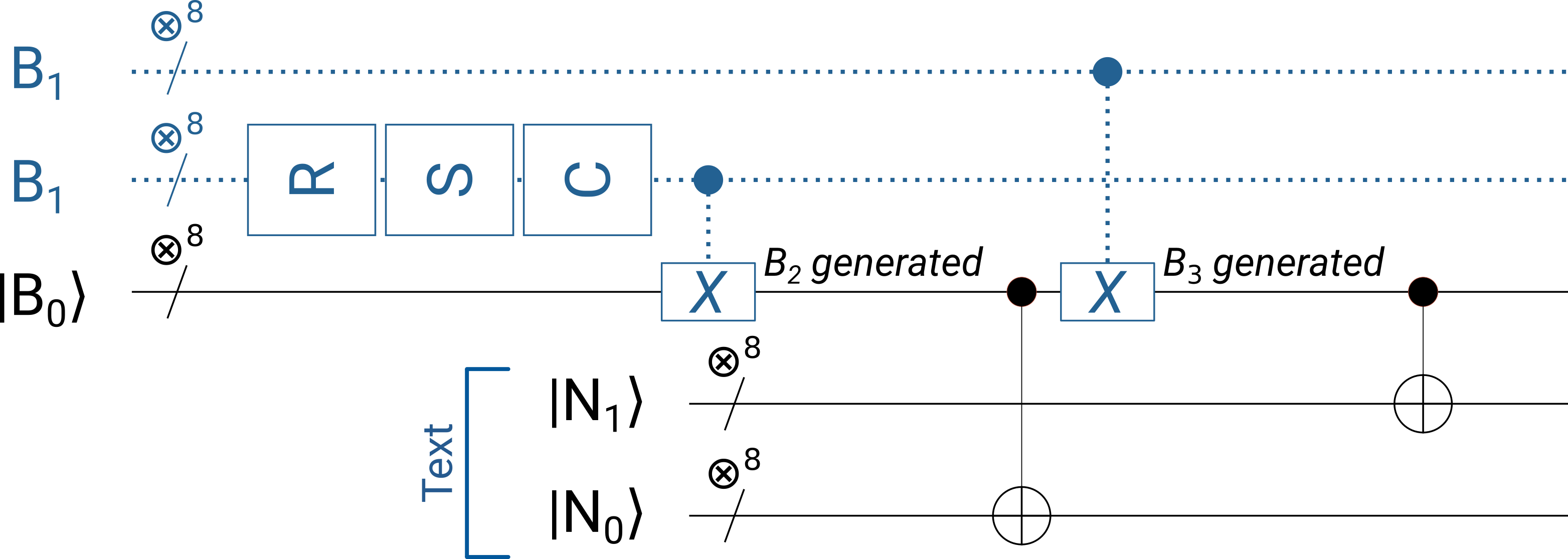}
	\caption{Quantum circuit for sequential $B_2$ and $B_3$ when $B_1$ is leaked. Key $B_0$ which is under Grover's attack and the text are stored in quantum memory. Classical data is shown with dotted line. Key register size is reduced from 8 to 4 qubits.}
	\label{fig:B3Gen}
\end{figure}

Following the diagram in figure \ref{fig:Round}, and keeping in mind that $B_1$ is stored in classical memory, we can easily add a key to the text by replacing the CNOT gate with a classically controlled Pauli-X gate, as well as generate $B_2$ by acting on $B_1$ with classical operations R, S and C, and then add it to $B_0$ again with classically-controlled X. Difficulties begin with $B_3$ generation. Without a leakage, CNOT with target qubits of $B_1$  is used for that. But with a leakage, we want to avoid storing $B_1$ in quantum memory, and the influence of quantum data on classical data can be provided only by a measurement, which will destroy the superposition generated in the key register by the Grower's algorithm and break the program. According to \ref{eq:keys} $B_3$ includes quantum data from $B_0$ and, as a result, cannot be stored in classical bits. However, this obstacle can be easily bypassed by sequential generation of keys $B_2$ and $B_3$ directly in $B_0$ register as shown on figure \ref{fig:B3Gen}. 

\subsection{The X S-box[X] problem}

The next move is to apply the second round subroutine quantum circuits to the text qubits as shown on figure (\ref{fig:Oracle}) and to generate final keys $B_4$ and $B_5$. Since we already have $B_3$ generated in quantum register, it is not difficult to generate $B_5$. According to \ref{eq:keys}, $B_2$ is reduced in $B_5$ generation, so it requires only quantum S and R operations, as well as xor with $C_1$ and $B_1$, which are classical data. 

\begin{figure}
	\centering
	\includegraphics[width=0.7\linewidth]{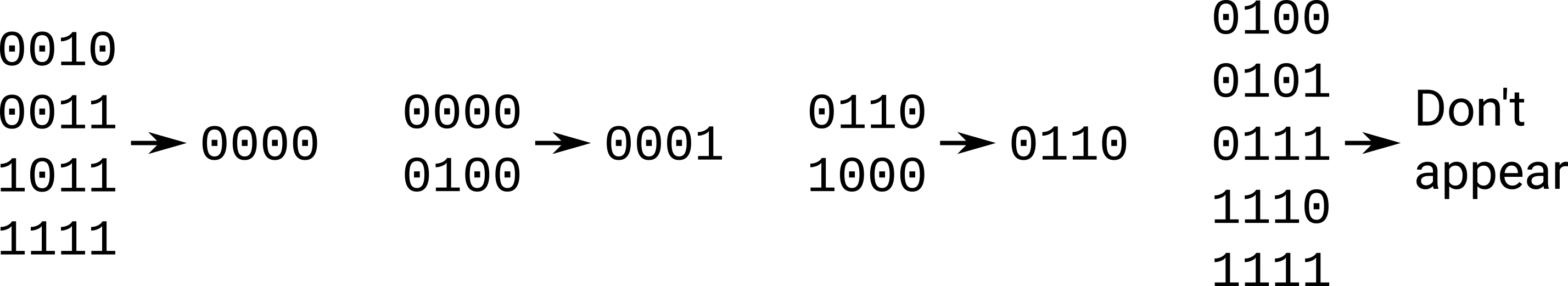}
	\caption{Collisions of the $X \oplus Sbox[X]$ transformation, that make it irreversible.}
	\label{fig:Collisions}
\end{figure}

Generation of $B_4$ is more complex. Following expression \ref{eq:keys}, $B_2$ can be expressed as
\begin{equation}
B_4 = C B_0 Sbox[C' B_0]
\label{B_4}
\end{equation}
where $C$ and $C'$ are some classical data. Basically here emerges the $X \oplus Sbox[X]$ transformation, which appears to be non unitary. It is well known that a quantum computer can only perform reversible unitary transformations. But the $X \oplus Sbox[X]$ transformation has one fourth order collision and two second order collisions, see figure \ref{fig:Collisions}.

To unravel the fourth-order collision for $0000$ at least 2 extra qubits are required. But the new 6-qubits transformation also needs to be unitary. The possible approach to find it is to manually construct the unitary matrix by replacing duplicate columns with columns that do not appear in the original transformation matrix. But then this six-qubit matrix will have to be decomposed into a product of elementary gates. The LIGHTER-R tool, that gave a convenient quantum circuit \ref{fig:Sbox} for the S-box is now developed only for 4-bit transformations \cite{Lighter-R}. And in general the elementary gate decomposition of a unitary matrix is a complex problem which solution leads to quantum circuits with exponential depth in the number of qubits \cite{Decomp}. For any 6-qubit unitary transformation that acts on 4-qubit subspace like $X \oplus Sbox[X]$ we haven't found a decomposition with acceptable circuit depth. 

\begin{figure}
	\centering
	\includegraphics[width=0.7\linewidth]{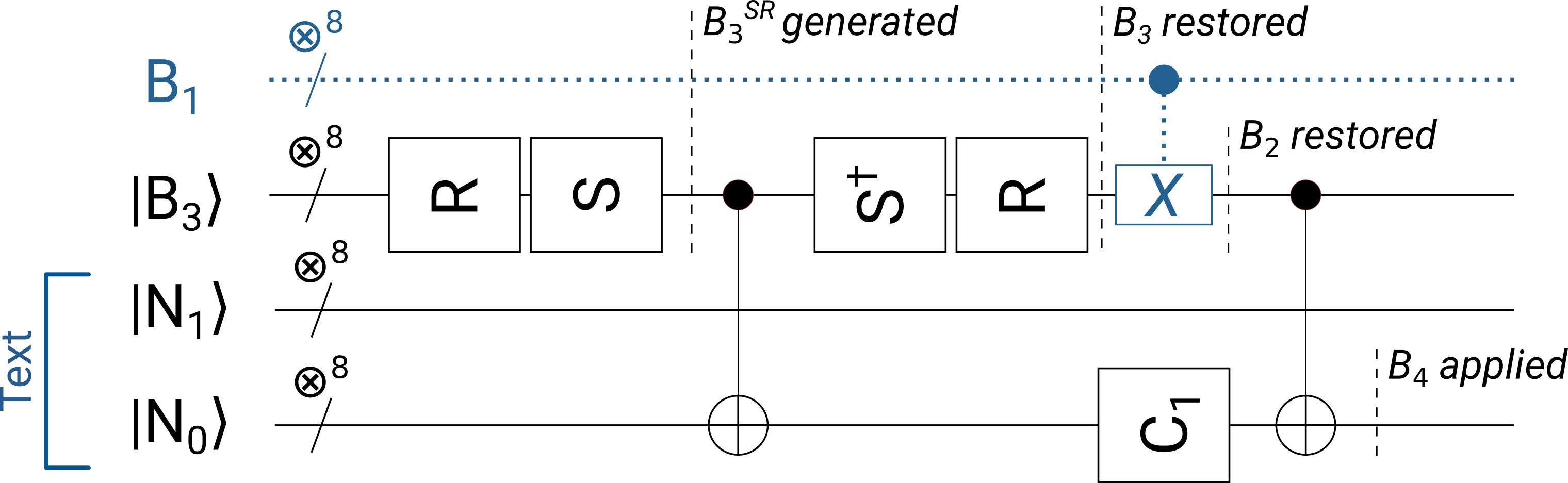}
	\caption{$B_4$ attack in case of $B_1$ leak. $B_4$ generation by itself is ununitary, but it's possible to generate $B_4$ sequentially and apply to the text.}
	\label{fig:B_4_Gen}
\end{figure}

However, for the oracle it's not actually necessary to generate $B_4$. The problem is to generate ciphertext for given key. And despite the nonunitarity of $B_4$ generation, due to xor additivity it is possible to generate the text with $B_4$ applied if text qubits are used. For this we can sequentially generate $C B_0$ from \ref{B_4}, apply it to the text, then return it to the $B_0$ state, add $C'$, apply S-Box and add it to the text again. More precisely, getting back to (\ref{eq:keys}), at the end of the first round key register contains $B_3$, so we better to compute $B_3^{SR}$, apply it to the text, then restore $B_2$ and also apply it to the text with $C_1$ as shown on figure \ref{fig:B_4_Gen}.

Therefore, the full algorithm of optimized oracle for $B_1$ leak is following, as shown in table \ref{table:B_1_Leak}. 

\begin{table}[H]
\begin{center}
	\begin{tabular}{||c| c c | c c c c||} 
		\hline
		Action & \multicolumn{2}{|c|}{Key} & \multicolumn{4}{c|}{Text}  \\ 
		\hline\hline
		Init & \bs{$B_0$} & $B_1$ & $N_0$& $N_1$ & $N_2$ & $N_3$ \\ 
		\hline
		Add key & \bs{$B_0$} & $B_1$ & \bs{$B_0 N_0$} & \bs{$B_0 N_1$} & $B_1 N_2$ & $B_1 N_3$ \\
		\hline
		\multicolumn{7}{|c|}{Round 1} \\
		\hline
		S & \bs{$B_0$} & $B_1$ & \bs{${B_0 N_0}^S$} & \bs{${B_0 N_1}^S$} & ${B_1 N_2}^S$ & ${B_1 N_3}^S$ \\
		\hline
		SR & \bs{$B_0$} & $B_1$ & \bs{${B_0 N_0}^S$} & ${B_1 N_3}^S$ & ${B_1 N_2}^S$ & \bs${B_0 N_1}^S$ \\
		\hline
		MC & \bs{$B_0$} & $B_1$ & \bs{$N_0^*$} & \bs{$N_1^*$} & \bs{$N_2^*$} & \bs{$N_3^*$} \\
		\hline
		\multirow {2}{4em}{Add key} & \bs{$B_2$} & $B_1$ & \bs{$B_2 N_0^*$} & \bs{$B_2 N_1^*$} & \bs{$N_2^*$} & \bs{$N_3^*$} \\
		& \bs{$B_3$} & $B_1$ & \bs{$B_2 N_0^*$} & \bs{$B_2 N_1^*$} & \bs{$B_3 N_2^*$} & \bs{$B_3 N_3^*$} \\
		\hline	
		\multicolumn{7}{|c|}{Round 2} \\
		\hline
		S & \bs{$B_3$} & $B_1$ & \bs$N_0^{' S}$ & \bs$N_1^{' S}$ &  \bs$N_2^{' S}$ &  \bs$N_3^{' S}$ \\
		\hline
		SR & \bs{$B_3$} & $B_1$ & \bs$N_0^{' S}$ & \bs$N_3^{' S}$ &  \bs$N_2^{' S}$ &  \bs$N_1^{' S}$ \\
		\hline
		\multirow {2}{4em}{Add key} & \bs{$B_3^{RSC}$} & $B_1$ & \bs{$B_3^{RSC} N_0^{**}$} & \bs{$B_3^{RSC} N_1^{**}$} & \bs{$B_3^{RSC} N_2^{**}$} & \bs{$B_3^{RSC} N_3^{**}$} \\
		& \bs{$B_2$} & $B_1$ & \bs{$B_4 N_0^{**}$} & \bs{$B_4 N_1^{**}$} & \bs{$B_5 N_2^{**}$} & \bs{$B_5 N_3^{**}$} \\
		\hline
	\end{tabular}
\caption{\label{table:B_1_Leak}Exact attack with $B_1$ leaked. 24 qubits are required}
\end{center}
\end{table}

This table shows states of key and qubit register step by step. Here and bellow bold shows quantum data, $N^*$ is used to mark text nibbles states before Add Key, $N^{'}$ and $N^{''}$ are used to mark resulting round nibbles after Add Key. The xor between an 8-bit round key $B_n$ and a 4-bit text nibble $N_m$ always implies the preservation of bit correspondence, i.e. xor of $B_0$ and $N_1$ means the bitwise xor of the nibble $N_1$ with the second nibble of $B_0$. This and the following tables omit elementary intermediate steps needed to get a new key applied to the text if all elements of a new key are stored in a register. For details, see (\ref{eq:keys}).

\subsection{$\boldsymbol{B_0}$-leak 25-qubits split attack}

Unfortunately, this algorithm is not universal. In case of $B_0$ leak, the X S-box[X] problem appears for $B_3$. At the same time $B_4$ generation requires to apply S-box for $B_3$. For that we need to have $B_3$ in a clear separate register, which excludes the possibility of its sequential generation. The question of how to generate S-box[$B_3$] if we know S-box[$N_2^* B_3$] is not considered in this paper. 

\begin{figure*}
	\centering
	\subfloat[$MC_0^0$]
	{
		\includegraphics[width=0.24\linewidth]{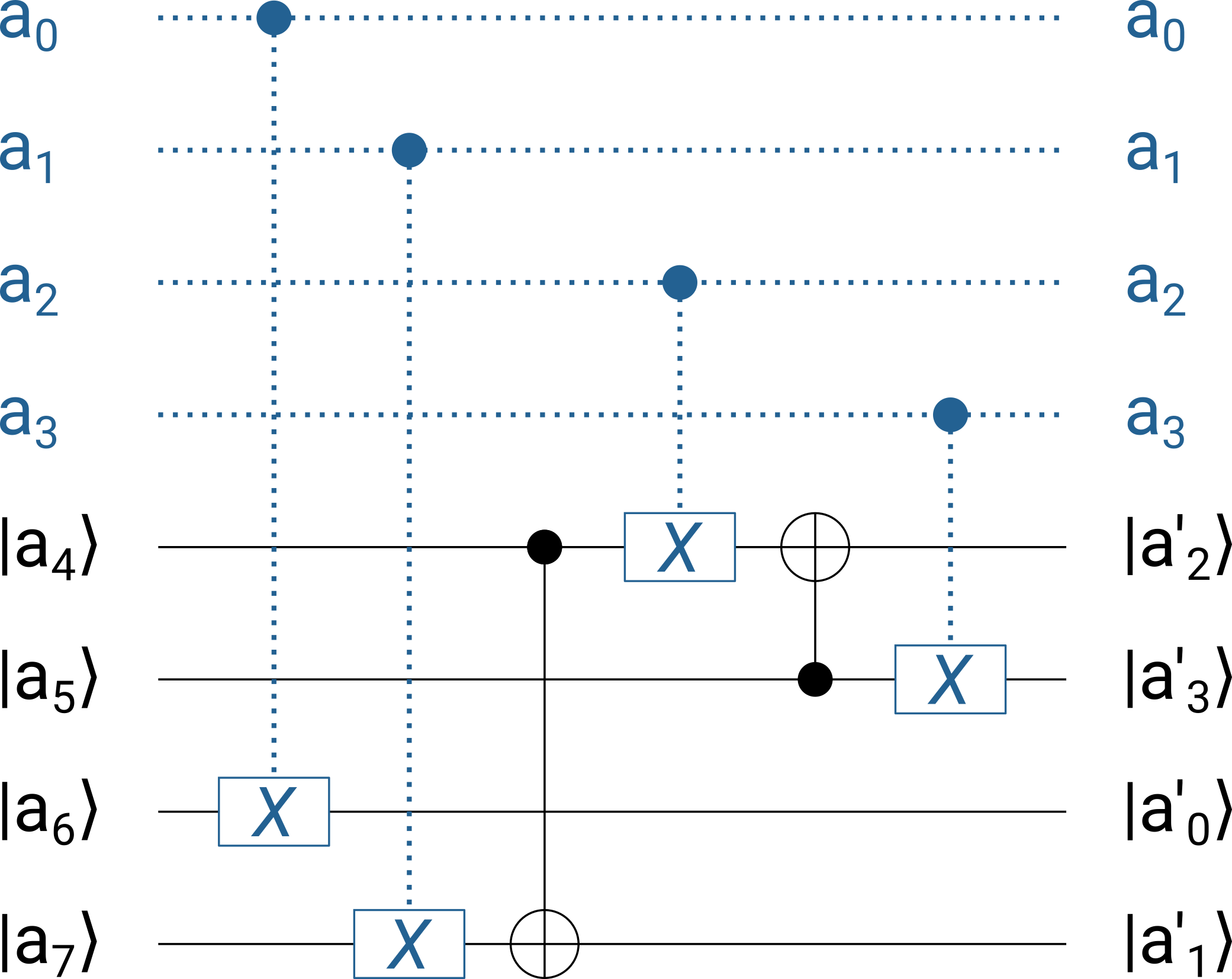}
		\label{fig:MC_0_0}
	}
	\subfloat[$MC_0^1$]
	{
		\includegraphics[width=0.24\linewidth]{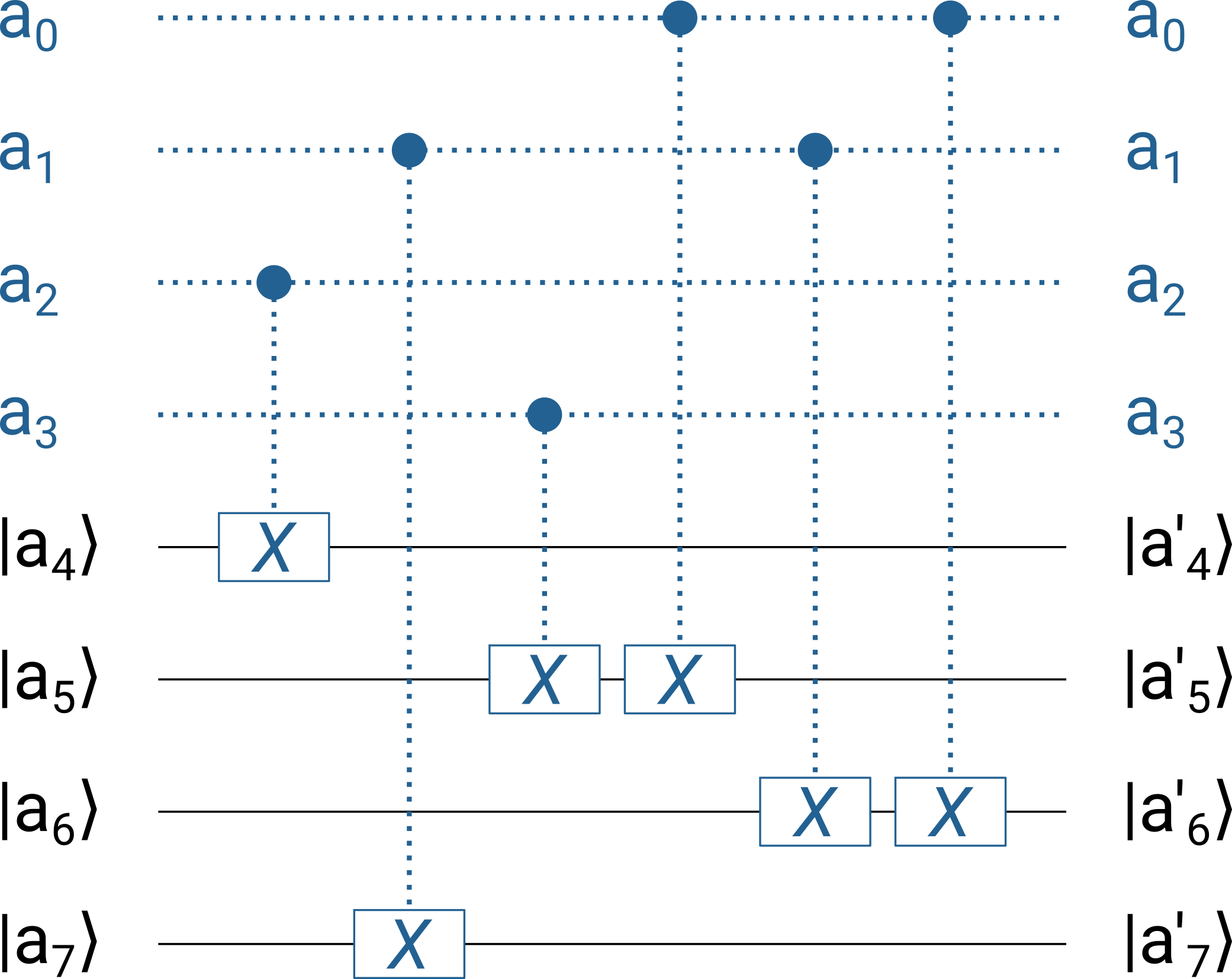}
		\label{fig:MC_0_1}
	}
	\subfloat[$MC_1^0$]
	{
		\includegraphics[width=0.24\linewidth]{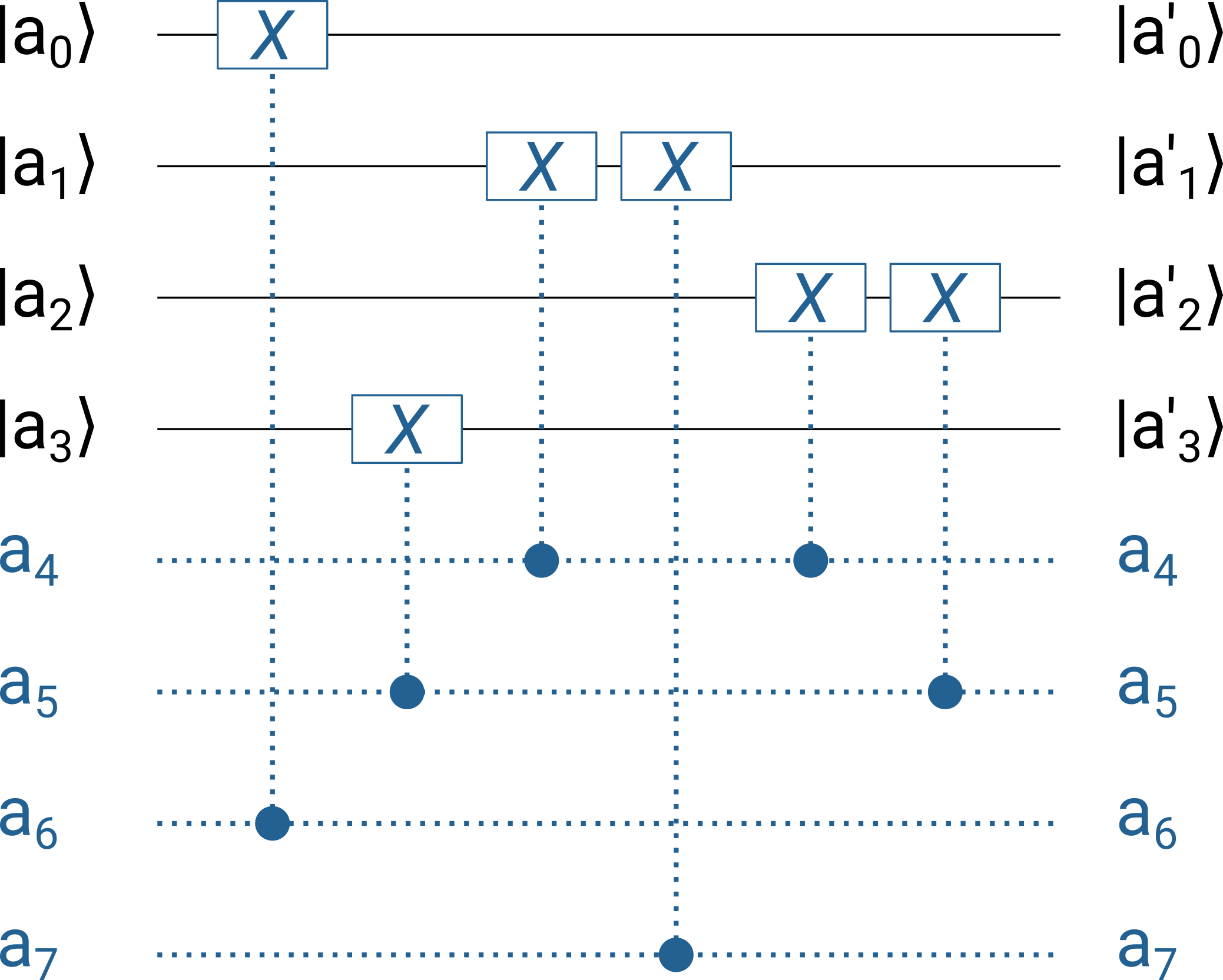}
		\label{fig:MC_1_0}
	}
	\subfloat[$MC_1^1$]
	{
		\includegraphics[width=0.24\linewidth]{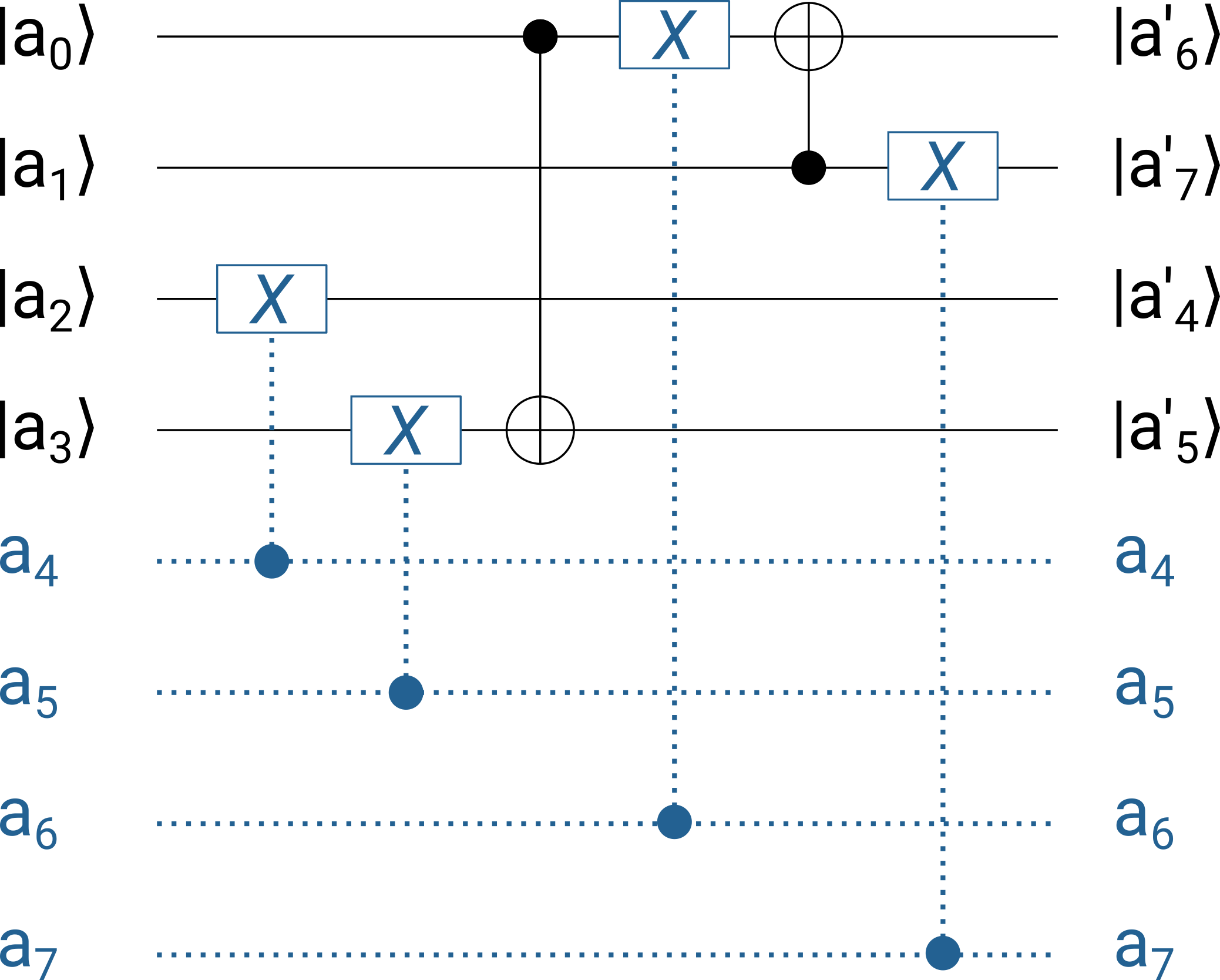}
		\label{fig:MC_1_1}
	}
	
	\caption{Possible optimized Mix Columns quantum circuits for split attack. The subscript indicates the number of the leaked nibble, the superscript indicates which nibble of the MC result is being generated}
	\label{fig:MC_Split}
\end{figure*}

Requiring 4 extra qubits to save pure $B_3$ would eliminate the advantage of the algorithm. However, we can use another approach that allows to save qubits. From table \ref{table:B_1_Leak} and (\ref{eq:MC}) you can see that Mix Column subroutine includes interaction between not only qubits but nibbles. In case of leakage, if part of Mic Column arguments are classical bits, as long as the result of this subroutine is defined like 8 qubits, it lead to a transition from classical to quantum data and the number of qubits increases. This can be avoided by generating results of the operation sequentially in the same qubit set. You can easily see from (\ref{eq:MC}) that it is not difficult to generate the first nor second nibble of Mix Columns result if one of the input nibbles is stored in classical data. In this way one $MC$ operation, which is a map on the space of 8 qubits, is replaced by four: $MC_{0, 1}^{0, 1}$, which are maps on the space of 4 qubits and take 4 more bits of classical information as parameters. Subscript shows, which part of the result byte is generated, when the superscript shows, which part of input byte is taken as classical data.
\begin{equation}
	\begin{aligned}
		\hat{MC}\ket{N_0, N_1} & = \hat{MC_0^0}(N_0)\ket{N_1} \otimes \hat{MC_1^0}(N_0)\ket{N_1} = \\ & = \hat{MC_0^1}(N_1)\ket{N_0} \otimes \hat{MC_1^1}(N_1)\ket{N_0}
	\end{aligned}
	\label{eq:MC_optimized_action}
\end{equation}
Here $\otimes$ stands for Kronecker product or qubit register concatenation. With this approach we are able to split whole second round to compute the first and then the other pair of the ciphertext nibbles in a single set of qubits. Possible quantum circuits for $MC_n^m$ are given at figure \ref{fig:MC_Split}. 

A complete algorithm of split attack in case of $B_0$ leakage is described step by step in table \ref{table:B_0_Leak_Split}. Rounds 1 and 2.1 are aimed to generate the first and the second nibbles of the ciphertext. For that, operations $MC_0^0$ and $MC_1^1$ are used. The generated nibbles are then compared with the ciphertext using a classically-controlled X gates and then multi-controlled CNOT-gate. The result $X_0$ of the comparison is written to an ancilla qubit, which is target for CNOT.
\begin{equation}
	\begin{aligned}
		\ket{N_0^{''}\oplus Ctxt_0, N_1^{''}\oplus Ctxt_1,X_0} &= \hat {CNOT}\ket{N_0^{''} \oplus Ctxt_0, N_1^{''} \oplus Ctxt_1}^{control}\ket{0}^{target} \\
		N_0^{''} &\equiv B_4 N_0^{'S}, N_1^{''} \equiv B_4 N_3^{'S}
	\end{aligned}
	\label{B_0_leak_ancilla}
\end{equation}
Here $\oplus$ stands for xor, $Ctxt_n$ represents n-th nibble of the ciphertext.
$X_0$ generation is followed by the data recovery procedure, which consists in the successive application of inverse transformations until the generation of $MC$ arguments. At this point operations  $MC_0^1$ and $MC_1^0$ are used to generate the other pair of ciphertext nibbles. The data obtained after round 2.2 can be xored with the ciphertext using classically controlled X-gates. Finally, multi-controlled CZ gate for 8 text qubits and one ancilla qubit $X_0$ will flip the phase of register state if initial keys $B_0$ and $B_1$ are correct. Thus the oracle functionality is achieved. 
\begin{equation}
	\hat U_{Oracle}\ket{B_0, B_1} = \hat {CZ}\ket{B_4 N_2^{'S} \oplus Ctxt_2, B_5 N_1^{'S} \oplus Ctxt_3, X_0}
	\label{B_0_leak_oracle}
\end{equation}

\begin{table}[H]
	\begin{center}
		\begin{tabular}{||c| c | c c c c| c||} 
			\hline
			Action & Key & \multicolumn{4}{c|}{Text} & Ancilla \\ 
			\hline\hline
			Init & $B_0$, \bs$B_1$ & $N_0$& $N_1$ & $N_2$ & $N_3$ & \\ 
			\hline
			Add key & $B_0$, \bs$B_1$ & $B_0 N_0$ & $B_0 N_1$ & \bs$B_1 N_2$ & \bs$B_1 N_3$ & \\
			\hline
			\multicolumn{7}{|c|}{Round 1} \\
			\hline
			S, SR & $B_0$, \bs$B_1$ & ${B_0 N_0}^S$ & \bs${B_1 N_3}^S$ & \bs${B_1 N_2}^S$ & ${B_0 N_1}^S$ & \\
			\hline
			$MC_0^0$, $MC_1^1$ & $B_0$, \bs$B_1$ &  \multicolumn{2}{c}{\bs$N_0^*$} & \multicolumn{2}{c|}{\bs$N_3^*$} & \\
			\hline
			Add key & $B_0$, \bs$B_2$, \bs$B_3$ & \multicolumn{2}{c}{\bs$B_2 N_0^*$} & \multicolumn{2}{c|}{\bs$B_3 N_3^*$} & \\
			\hline	
			\multicolumn{7}{|c|}{Round 2.1} \\
			\hline
			S, SR & $B_0$, \bs$B_2$, \bs$B_3^{RSC}$ & \bs$N_0^{'S}$ & \bs$N_3^{' S}$ & - &  - & \\
			\hline
			Add key & $B_0$, \bs$B_4$ \bs{$B_3^{RSC}$} & \bs{$B_4 N_0^{'S}$} & \bs{$B_4 N_3^{'S}$} & - & - & \bs$X_0$\\ 
			\hline			
			\multicolumn{7}{|c|}{Data recovery} \\
			\hline
			Add key${}^\dagger$ & $B_0$, \bs$B_4$ \bs{$B_3^{RSC}$} & \bs{$N_0^{'S}$} & \bs{$N_1^{'S}$} & - & - & \bs$X_0$\\ 
			\hline
			SR${}^\dagger$, S${}^\dagger$ & $B_0$, \bs$B_2$, \bs$B_3^{RSC}$ & \bs$N_0^{'}$ & - & \bs$N_3^{'}$ & - & \bs$X_0$ \\
			\hline
			Add key${}^\dagger$ & $B_0$, \bs$B_2$, \bs$B_3$ & \multicolumn{2}{c}{\bs$N_0^*$} & \multicolumn{2}{c|}{\bs$N_3^*$} & \bs$X_0$ \\
			\hline
			$MC_0^{0\dagger}$, $MC_1^{1\dagger}$ & $B_0$, \bs$B_2$, \bs{$B_3$} & ${B_0 N_0}^S$ & \bs${B_1 N_3}^S$ & \bs${B_1 N_2}^S$ & ${B_0 N_1}^S$ & \\
			\hline
			\multicolumn{7}{|c|}{Round 2.2} \\		
			\hline	
			$MC_0^1$, $MC_1^0$ & $B_0$, \bs$B_2$, \bs$B_3$ &  \multicolumn{2}{c}{\bs$N_1^*$} & \multicolumn{2}{c|}{\bs$N_2^*$} & \bs$X_0$\\
			\hline
			Add key & $B_0$, \bs$B_2$, \bs$B_3$ & \multicolumn{2}{c}{\bs$B_2 N_1^*$} & \multicolumn{2}{c|}{\bs$B_3 N_2^*$} & \bs$X_0$\\			
			\hline
			S, SR & $B_0$, \bs$B_2$, \bs$B_3^{RSC}$ & - & - & \bs$N_2^{'S}$ & \bs$N_1^{' S}$  & \bs$X_0$\\
			\hline
			Add key & $B_0$, \bs$B_4$ \bs$B_3$ & - & - & \bs{$B_4 N_2^{'S}$} & \bs{$B_5 N_1^{'S}$}  & \bs$X_0$\\
			\hline
		\end{tabular}
		\caption{\label{table:B_0_Leak_Split}Split attack with $B_0$ leaked. 25 qubits are required. 17 qubits are required in case of $B_1$ leakage}
	\end{center}
\end{table}

This approach, when combined with the key expansion technique described in the previous section, provides a 17-qubit attack in the case of a $B_1$ leak. But extending this approach to generate nibbles of ciphertext one by one allows for even greater savings in qubits, and, more importantly, to construct a more universal attack mechanism.

\subsection{Double split attack}

It is obvious that the circuits on figure \ref{fig:MC_Split}, by replacing classically controlled X-gates with CNOT gates, can be turned to fully quantum ones. Then we get a set of circuits, the input of which contains 8 qubits of the MC argument, and the output contains the first or second nibble of the MC result, as well as the first or the second input nibble unchanged. That allows to use qubits of the key register in MC procedure without affecting one of the input key nibbles. Note that in the case of fully quantum inputs, there is no practical difference between $MC_0^1$ and $MC_1^1$, as well as between $MC_0^0$ and $MC_1^0$.  

Also, this approach allows to handle one-nibble key leaks. Therefore, further the keys will be denoted by nibbles: subscript indicates the number of key, superscript - the number of nibble, $B_0 = [B_0^0, B_0^1]$, $B_1 = [B_1^0$, $B_1^1]$. And since the Rotate $(R)$ operation in the key expansion just rearranges the nibbles, its effect is reduced to changing the order of elements: $(B_1)^R = [B_1^0, B_1^1]^R = [B_1^1, B_1^0]$. 

A complete algorithm of double split attack in case of $B_0^0$ leakage is described step by step in table \ref{table:Double_Split}. The description of data recovery procedure is omitted due to the triviality of performing the inverse of unitary transformations chain. AddTxt operation is an equivalent of Add Key, but the name notes, that in this case classical text is added to the quantum key register. Obviously, AddTxt is self-inverse operation. 8 subrounds are required to generate 4 nibbles of ciphertext. Therefore algorithm needs 3 ancilla qubits $X_0$, $X_2$ and $X_3$ to store the results of comparing the generated nibbles with corresponding ciphertext nibbles. 

\begin{table}
	\begin{center}
		\begin{tabular}{||c|c|c|c||} 
			\hline
			Action & Key & Text & Ancilla \\ 
			\hline\hline
			Init & $B_0^0$, \bs$B_0^1$, \bs$B_1^0$, \bs$B_1^1$ & $N_0$, $N_1$, $N_2$, $N_3$ & \\ 
			\hline
			\multicolumn{4}{|c|}{Round 1.1} \\
			\hline
			AddTxt, S, SR & \multicolumn{2}{c|}{$\underline{(B_0^0 N_0)^S, \pmb{(B_1^1 N_3)^S}}$, \bs$B_0^1$, \bs$B_1^0$} & \\
			\hline
			$MC_0^0$, $S^\dagger$, AddTxt & $B_0^0$, \bs$B_0^1$, \bs$B_1^0$, \bs$B_1^1$ & \bs$N_0^*$ & \\
			\hline 
			Add key & $B_0^0$, \bs$B_0^1$, \bs$B_1^0$, \bs$(B_1^1)^{SC}$ & \bs$N_0' = B_2^0 N_0^*= B_0^0 (B_1^1)^{SC} N_0^*$ & \\
			\hline
			\multicolumn{4}{|c|}{Round 2.1} \\
			\hline
			\multirow {2}{*}{S, SR, AddKey} & \multirow {2}{*}{$B_0^0$, \bs$B_0^1$, \bs$B_1^0$, \bs$B_1^1$} & \bs$N_0^{''} = B_4^0 (N_0^{'})^{S} = $ & \multirow{2}{*}{\bs$X_0$} \\ && \bs$= B_0^0 (B_1^1)^{SC} (B_1^1 B_0^1 (B_1^0)^{SC})^{SC}(N_0^{'})^{S}$ & \\
			\hline
			
			\multicolumn{4}{|c|}{Round 1.2} \\
			\hline
			Data recovery & \multicolumn{2}{c|}{$\underline{(B_0^0 N_0)^S, \pmb{(B_1^1 N_3)^S}}$, \bs$B_0^1$, \bs$B_1^0$} & \bs$X_0$ \\
			\hline
			$MC_0^1$, $S^\dagger$, AddTxt & $B_0^0$, \bs$B_0^1$, \bs$B_1^0$, \bs$B_1^1$ & \bs$N_1^*$ & \bs$X_0$ \\
			\hline 
			Add key & $B_0^0$, \bs$B_0^1$, \bs$(B_1^0)^{SC}$, \bs$B_1^1$ & \bs$N_1' = B_2^1 N_1^*= B_0^1 (B_1^0)^{SC} N_0^*$ & \bs$X_0$ \\
			\hline
			
			\multicolumn{4}{|c|}{Round 2.2} \\
			\hline
			\multirow {2}{*}{S, SR, AddKey} & \multirow {2}{*}{$B_0^0$, \bs$B_0^1$, \bs$B_1^0$, \bs$B_1^1$} & \bs$N_3^{''} = B_5^1 (N_1^{'})^{S} = $ & \multirow{2}{*}{\bs$X_0, X_3$} \\ && \bs$= B_1^1 (B_1^0 B_1^1 B_0^0 (B_1^1)^{SC})^{SC}(N_1^{'})^{S}$ & \\
			\hline
			
			\multicolumn{4}{|c|}{Round 1.3} \\
			\hline
			Data recovery, & \multicolumn{2}{c|}{\multirow {2}{*}{$B_0^0$, \bs$B_1^1$, \bs$\underline{(B_1^0 N_2)^S, (B_0^1 N_1)^S}$}} & \multirow {2}{*}{\bs$X_0, X_3$} \\
			AddTxt, S, SR &\multicolumn{2}{c|}{} & \\
			\hline	
			$MC_0^0$, $S^\dagger$, AddTxt & $B_0^0$, \bs$B_0^1$, \bs$B_1^0$, \bs$B_1^1$ & \bs$N_2^*$ & \bs$X_0, X_3$ \\
			\hline 
			Add key & $B_0^0$, \bs$B_0^1$, \bs$B_1^0$, \bs$(B_1^1)^{SC}$ & \bs$N_2' = B_3^0 N_2^*= B_1^0 B_0^0 (B_1^1)^{SC} N_2^*$ & \bs$X_0, X_3$ \\
			\hline
			
			\multicolumn{4}{|c|}{Round 2.3} \\
			\hline
			\multirow {2}{*}{S, SR, AddKey} & \multirow {2}{*}{$B_0^0$, \bs$B_0^1$, \bs$B_1^0$, \bs$B_1^1$} & \bs$N_2^{''} = B_5^0 (N_2^{'})^{S} = B_1^0 (B_3^1)^{SC} (N_2^{'})^{S} = $ & \bs$X_0, X_2$ \\ && \bs$= B_1^0 (B_1^1 B_0^1 (B_1^0)^{SC})^{SC}(N_2^{'})^{S}$ & \bs$X_3$\\
			\hline
			
			\multicolumn{4}{|c|}{Round 1.4} \\
			\hline
			\multirow {2}{*}{Data recovery} & \multicolumn{2}{c|}{\multirow {2}{*}{$B_0^0$, \bs$B_1^1$, \bs$\underline{(B_1^0 N_2)^S, (B_0^1 N_1)^S}$}} & \bs$X_0, X_2$ \\
			 &\multicolumn{2}{c|}{} & \bs$X_3$ \\
			\hline	
			$MC_0^1$, $S^\dagger$, & 	\multirow {2}{*}{$B_0^0$, \bs$B_0^1$, \bs$B_1^0$, \bs$B_1^1$ }& 	\multirow {2}{*}{\bs$N_3^*$} & \bs$X_0, X_2$ \\ 
			AddTxt & & & \bs$X_3$ \\
			\hline 
			\multirow {2}{*}{Add key} & \multirow {2}{*}{$B_0^0$, \bs$B_0^1$, \bs$(B_1^0)^{SC}$, \bs$B_1^1$} & \multirow {2}{*}{\bs$N_3' = B_3^1 N_3^*= B_1^1 B_0^1 (B_1^0)^{SC} N_3^*$} & \bs$X_0, X_2$ \\ &&& \bs$X_3$ \\
			\hline
		
			\multicolumn{4}{|c|}{Round 2.4} \\
			\hline
			\multirow {2}{*}{S, SR, AddKey} & \multirow {2}{*}{$B_0^0$, \bs$B_0^1$, \bs$B_1^0$, \bs$B_1^1$} & \bs$N_1^{''} = B_4^0 (N_3^{'})^{S} = B_2^0 (B_3^1)^{SC} (N_1^{'})^{S} = $ & \bs$X_0, X_2$ \\ && \bs$= B_2^0 (B_1^1 B_0^1 (B_1^0)^{SC})^{SC}(N_1^{'})^{S}$ & \bs$X_3$\\
			\hline
			
		\end{tabular}
		\caption{\label{table:Double_Split}Double Split attack with $B_0^0$ leaked. Underlined pairs of nibbles are subjected to one the optimized MC transformation from figure \ref{fig:MC_Split} on the next step. As a result, one nibble saves the result of the transformation, and the second remains unchanged. Then, using the inverse transformation $S^\dagger$ and re-addition with the corresponding text $N$ it's possible to restore the original value of the key $B$ in that nibble. }
	\end{center}
\end{table}

Similar to the previous section, ancilla qubits generation and oracle action are given by
\begin{equation}
	\begin{aligned}
		\ket{N_n^{''} \oplus Ctxt_n}\ket{X_n} &= \hat {CNOT}\ket{N_n^{''} \oplus Ctxt_n}^{control}\ket{0}^{target}\\
		\hat U_{Oracle}\ket{B_0, B_1} &= \hat {CZ}(X_0, \ket{N_1^{''}} \oplus Ctxt_1, X_2, X_3)
	\end{aligned}
\end{equation}

A major advantage of double split approach is its versatility. Since the keys are stored unchanged during almost the entire algorithm, and are restored after the MC operation, it is possible to attack any nibble leak configuration. 19 qubits are required for any 1 nibble leakage. 15 and 11 qubits are required for of 2 and 3 nibble leaks, respectively. Moreover, in the absence of a leak, this algorithm allows to run attack with 23 qubits, while the best previously known oracle required 32 qubits.

\section{Discussion}

\subsection{Results}
The main result of this work is the construction of S-AES attack algorithms that can be effectively simulated. While the previously known algorithm was able to carry out an attack only on the full key, which required 32 qubits and excluded the possibility of simulating this quantum algorithm on the GPU without tensor networks with strongly limited accuracy, in this work we found an algorithm capable of performing the same attack with only 23 qubits. Also, this algorithm is capable of performing an attack using 19, 17 15 or 11 qubits in case of a key leak. A simpler algorithm that performs an attack with 24 qubits in case of $B_1$ key leak was simulated in the MSU Remote Quantum Computing system. In addition to new prospects for modeling practically significant error mitigation methods, the demonstrated approach opens up a new direction of research in the field of quantum cryptanalysis, demonstrating that a quantum computer can be used not only as a full-stack cracker, but also as one of the many attack tools. It also makes possible to greatly reduce the requirements for the number of qubits that a quantum computer must have for potential use in cryptanalysis, which probably exacerbates the problem of quantum threat and makes the prospect of using NISQ-devices in cryptanalysis more real.

\subsection{Problems}
The major problem of all described algorithms is that the sequential generation of ciphertext elements multiply increases the depth of the quantum circuit. In the case of a double split attack without leakage, a gain of about a third of qubits is accompanied by an almost fourfold increase in circuit depth, which clearly negatively affects the possibilities of experimental implementation of this algorithm. On the other hand, the new approaches are significantly more flexible than Grover's full attack with a basic oracle. The use of split and double split attacks in combination with error mitigation and error correction methods allows us to talk about the possibility of experimental implementation of an attack on S-AES with leakage on logical qubits even using NISQ devices. Today's largest quantum computer with 433 qubits \cite{IBM} in the context of the attack on 1 nibble of a key that requires 11 qubits, allows us to talk about the ratio of 39 physical qubits to one logical.

Another problem is the construction of multi-control multi-qubit gates. In the case of an elementary attack on the full text, see figure \ref{fig:Iteration}, when applying multi-qubit gates, the register always contains qubits that are not involved in the operation, and then can be used as borrowed ones. A well-known algorithm allows one to construct a multi-control CNOT or CZ gates from three-qubit Toffoli gates with a linear circuit depth using borrowed qubits \cite{Elementary_gates}. An algorithm for such decomposition without ancilla qubits also exists, but requires two-qubit gates more complex than CNOT and CZ \cite{Toffoli}.

\subsection{Further research}
The developed algorithms provide new opportunities in the simulation of quantum attacks and the model study of error mitigation methods. The subject of further research will be the determination of the resistance of these algorithms to various types of realistic quantum errors, the development of specific error mitigation methods and, in the future, the experimental implementation of at least separate elements of the attack.

Another area of research could be the development of methods to reduce the depth of attack circuit. Elementary attack algorithms imply a complete reversal of the oracle action to recover the initial keys $B_0$ and $B_1$ before the action of the Grover operator. Generally speaking, this is not necessary, since the Grover operator only works on amplifying the amplitude of the state component with flipped phase. The oracle, in the process of preparing the ciphertext, works with qubits as with digital data, without changing the phase of state components. The component whose phase is affected by the correct oracle with CZ gate always remains corresponding to the correct solution of the search problem. Thus, the depth of attack can be almost halved if the recovery of original keys is removed in successive iterations of the Grover algorithm, and the oracles themselves are properly modified to work with the keys and ciphertexts generated at the previous iteration.The overhead costs in this case will be ancilla qubits $X_0$ $X_2$ and $X_3$. Clearing their states would require either a non-physical procedure for resetting an arbitrary qubit state to $\ket{0}$, or a significantly more complex algorithm.

Finally, the modular approach on which the double split attack is built gives the algorithm good scalability. 
In the context that double split attack conceptually requires only qubits to store key bits subjected to the Grover's attack and a set of qubits to store one elementary piece of text, the prospect of scaling the algorithm to full AES looks very promising. In the future, the possibility of extending this approach to a full AES will be studied and then elements of this attack will be simulated with realistic noise parameters.

%
%
\def\refname{References}

%
%
\end{document}